# Area-wide traffic signal control based on a deep graph Q-Network (DGQN) trained in an asynchronous manner

**Gyeongjun Kim**
Graduate student
Department of Urban Engineering, Chung-Ang University, Seoul, Korea
84 Heukseok-ro, Dongjak-gu, Seoul 156-756, Korea
kkjunn7033@cau.ac.kr

**Keemin Sohn*** Corresponding author
Professor
Department of Urban Engineering, Chung-Ang University, Seoul, Korea
84 Heukseok-ro, Dongjak-gu, Seoul 156-756, Korea
kmsohn@cau.ac.kr

## Abstract

Reinforcement learning (RL) algorithms have been widely applied in traffic signal studies. There are, however, several problems in jointly controlling traffic lights for a large transportation network. First, the action space exponentially explodes as the number of intersections to be jointly controlled increases. Although a multi-agent RL algorithm has been used to solve the curse of dimensionality, this neither guaranteed a global optimum, nor could it break the ties between joint actions. The problem was circumvented by revising the output structure of a deep Q-network (DQN) within the framework of a single-agent RL algorithm. Second, when mapping traffic states into an action value, it is difficult to consider spatio-temporal correlations over a large transportation network. A deep graph Q-network (DGQN) was devised to efficiently accommodate spatio-temporal dependencies on a large scale. Finally, training a RL model to jointly control traffic lights in a large transportation network requires much time to converge. An asynchronous update methodology was devised for a DGQN to quickly reach an optimal policy. Using these three remedies, a DGQN succeeded in jointly controlling the traffic lights in a large transportation network in Seoul. This approach outperformed other “state-of-the-art” RL algorithms as well as an actual fixed-signal operation.

## Keywords

Adaptive traffic signal control, Deep graph Q-network (DGQN), Graph convolution, Reinforcement learning,

## 1. Introduction

Many researchers have rigorously studied the application of reinforcement learning (RL) to traffic light control. A multi-agent implementation prevails because a single-agent RL algorithm cannot accommodate the increasing size of state and action spaces when network-wide traffic signals are jointly controlled. A multi-agent RL algorithm, however, cannot guarantee a global optimum without full coordination with other agents' actions, and it cannot break ties between agents when they cooperate to reach a common goal (Busoniu, 2010). Our previous study resolved these problems within the framework of a single-agent model by revising the output structure of the original deep Q-network (DQN) (Lee et al., 2019). Whereas the extended model had been validated for a small-scale synthesized network composed of only 4 intersections, in the present study the model was applied to jointly control a real transportation network containing 15 intersections. In addition, mathematical expressions of the extended DQN were newly established in a rigorous manner. An ideal RL algorithm must circumvent the curse of dimensionality with a single agent. The extended DQN takes the form of a single-agent model and is sufficient for a large action space. The number of output nodes of the extended DQN is linearly proportional to the number of intersections to be jointly controlled, whereas that of the original DQN exploded as the number of intersections grew larger.

The second problem is posed when a RL-based traffic signal control is implemented in a complex transportation network. It is inefficient for either a fully connected (FC) layer or a 2-D convolutional layer to recognize the topology of a transportation network. Thus, those methods require an inordinate amount of computing time to accommodate the spatial dependencies between the traffic states of different intersection approaches. A transportation network has a constant topology through which traffic states propagate. If a graph convolution is employed in the structure of a Q-network, the dependency of the traffic state of a given intersection on the traffic states of other intersections can be efficiently addressed when approximating the true action values. Some researchers have pioneered the use of graph convolutions to constitute a Q-network for traffic signal control (Wang et al., 2020; Nishi et al., 2018). Their test beds, however, included only a few intersections and depended on a multi-agent RL algorithm that could not fully coordinate the actions of different agents. We devised a novel graph convolution scheme that can vary the connection intensities and incorporate them into an extended DQN for joint traffic signal control. A deep graph Q-network (DGQN) was developed to accommodate the spatio-temporal dependencies in a transportation network. The DGQN made it possible to accommodate the traffic states of upstream and downstream roads from past times in order to more efficiently determine the present traffic signals of an intersection approach. It is also possible to use either a FC or a 2-D convolutional layer to recognize spatial correlations. A DGQN, however, is the most efficient model because it directly uses the topology of a road network to derive action values.

The difficulty in training a DGQN to jointly control traffic lights on a large scale was the last issue to be addressed in the present study. The proposed DGQN cannot reduce a large action space but can deal with the problem within the revised architecture of a DQN using limited computer resources. It should be noted that the large action space remains unchanged and should be explored sufficiently during training. For convergence, a DGQN must experience as many transitions as possible from an action-state pair to a subsequent state. In this context, a single environment is not sufficient for a RL algorithm to reach the convergence within a practical computation time. In the present study, multiple environments were set up and asynchronous updates were applied in training the DGQN. Unlike the existing A3C algorithm developed by Mnih et al. (2016), a replay memory was separately kept for each actor-learner. Four simulated environments were used, and each environment had a different level of traffic demand.

The proposed DGQN essentially took the form of a single-agent RL model rather than the multi-agent RL models that prevail in the field of study for joint traffic signal control at a network level. Nonetheless, owing to the three remedies outlined above the DGQN made it possible to find an optimal policy that could jointly control the traffic signals of 15 intersections in a real transportation network. The DGQN outperformed a DQN with only FC layers and an original graph convolutional neural network (GCN) with a constant adjacency matrix. Furthermore, an asynchronous learning scheme made it possible for a DGQN to converge to an optimal policy within a practical amount of computing time.

The next section introduces a literature search that dealt with the applications of a RL algorithm to traffic signal control. In the third section, the state, action, and reward are defined for the proposed DGQN. The fourth section describes the structure of the proposed DGQN to circumvent the curse of dimensionality with respect to a large action space. An asynchronous training algorithm is devised in the fifth section. How to set up simulation experiments is described in the sixth section. The seventh section shows analysis results and performance comparisons with other reference models. In the last section, conclusions are drawn, and further studies are discussed.

## 2. Related work

Studies that employed a RL-based algorithm to jointly control traffic signals in a transportation network are investigated in this section. Prior to the advance of deep learning technologies, model-based multi-agent RL algorithms had already been adopted in joint traffic signal control to gain an advantage of saving computing resources in a distributed manner (Wiering, 2000; Steingrover et al., 2005; Kuyer et al., 2008; Houli et al., 2010; Zhu et al., 2015; and El-Tantawy et al., 2013). Wiering (2000) established an RL model prototype for traffic signal control based on multiple agents. Later, Kuyer et al. (2008) extended Wiering's model by employing a Max-Plus algorithm to pass messages among neighboring agents, which facilitates coordination between actions of different agents. Houli et al. (2010) devised a vehicular ad hoc network within the framework of a multi-agent RL to streamline the information exchange among agents. More practically, a partially coordinated multi-agent RL algorithm was developed by El-Tantawy et al. (2013), which can be applied to real-world applications by allowing each agent to communicate only with its first-order neighbors and to partition the entire state space into subspaces consisting of a pair of agents. All these researchers, however, adopted a tabular update of the Q-function and did not approximate their Q-function using a deep neural network.

A model-free RL algorithm that approximates the Q-function using a deep neural network started to emerge for traffic light control after a DQN had succeeded in the playing of classic video games (Mnih et al., 2015). There are only a few references in the literature to the adoption of a deep-learning model to address issues associated with joint traffic signal control in a large transportation network. Some researchers used a DQN within a multi-agent RL framework to solve the joint traffic signal control problem (Van der Pol and Oliehoek, 2016; Tan et al., 2019). However, they did not ensure that an exact solution was found because only partial coordination between agents was considered. Van der Pol and Oliehoek (2016) decomposed the global Q-function into local Q-functions and used a Max-Plus algorithm to exchange messages between agents within a DQN framework. Tan et al. (2019) used a hierarchy to decompose a RL problem, and a global agent utilized the achievements from local agents to perform its own action. A policy-gradient RL algorithm was also adopted to solve joint traffic signal control in a large transportation network (Casas, 2017; Chu et al., 2019). Although Casas (2017) applied a single-agent actor-critic algorithm to find a global solution, he had to fix the sequence of traffic signal phases and simply optimized the ratio of phases with respect to the common signal cycle time. Chu et al. (2019) employed a multi-agent actor-critic algorithm that shares partial information between agents. They included information of neighborhood policies to improve the observability of each local agent and introduced a spatial discount factor to weaken the state and reward signals from other agents. A graph-based neural network is, however, more appropriate for such spatial interactions.

Some pioneers have begun to use a graph-based neural network to approximate the true Q-function when traffic lights are jointly controlled in a transportation network. When constructing a DQN, Nishi et al. (2018) used graph convolutional layers to accommodate spatial dependencies between traffic states separated geographically in a transportation network. An adjacency matrix that indicates the first-order connections in a road network was prepared in advance. In their framework, however, each traffic signal is controlled by a separately learned policy, which means they could not find a global joint policy. More recently, Wang et al. (2020) adopted an attention-based graph convolution to consider spatial dependencies between traffic states when approximating the true Q-function. The directional traffic light adjacency graph was explicitly constructed for modeling the geographical structure information to facilitate coordination among multi-intersection traffic light control. A recurrent neural network was also integrated with the graph units to accommodate temporal dependencies. Spatio-temporal correlations between traffic states were recognized simultaneously within a DQN framework. Their DQN was evaluated simply in a distributed manner whereby the global Q-function is decomposed into local Q-functions. As far as we could ascertain, there is no previous multi-agent RL algorithm for joint traffic light control that can include full coordination between the actions of other agents.

## 3. The definitions of state, action, and reward

### 3.1 The definition of state

Conventional traffic parameters such as delays and queue lengths are a decisive measure to evaluate the performance of an adaptive traffic signal control system. In the field of study for RL-based traffic signal controls, the traffic delay and queue length have widely been chosen to represent the traffic state (Abdoos at al., 2014; Arel et al., 2010; Bakker et al., 2010; El-Tantawy et al., 2013; and Jin and Ma, 2015; and Isa et al., 2006). The present study also adopted both of these parameters as the state of the traffic environment. However, there may be skepticism regarding how to accurately measure these parameters when an RL algorithm is implemented in the field. Our two previous studies showed strong evidence that they can be measured onsite based solely on video images (Shin et al., 2019; Chung and Sohn, 2017). The state was defined for each lane group that is comprised of one or more lanes that share a common stop-line and capacity. Generally, all exclusive turn lanes are treated as separate lane groups. Through lanes are also generally grouped together, including through lanes that allow for shared right and/or left-turn movements. Feasible phases and lane groups should be recognized for each intersection in a transportation network prior to establishing a DQN model.

### 3.2 The definition of action

There are two types of actions in a RL-based traffic signal control problem. A continuous action allows a phase duration or its proportion to a cycle length to be determined (Bakker et al., 2010; El-Tantawy et al., 2013; and Wiering, 2000). In this case, the displaying sequence of given phases had to be fixed. On the other hand, for a discrete action any feasible phase can be selected for a time period (Balaji et al., 2010; Abdoos et al., 2013; Abdoos et al., 2011; and Casas, 2017). For the discrete action, the duration of a signal phase cannot be shortened below the predefined interval but should be a multiple value of the interval. The present study chose the second option to secure a larger degree of freedom in a traffic signal control. Feasible phases for each intersection were selected relative to the real operation of a traffic signal control in the testbed. For the discrete action space, it should be noted that the number of joint actions explodes exponentially with increases in the number of intersections to be jointly controlled.

### 3.3 The definition of reward

The goal of joint traffic signal control is to minimize the total delay at the transportation network level. The conventional algorithm for adaptive traffic light control has not achieved this goal in the real world. Most adaptive signal control systems cannot help leading to a local optimum. Applying the best offset for a major corridor may undermine minor traffic flows, which is frequently observed in the field. Even though a multi-agent RL algorithm chooses a delay-dependent reward specification, it ends up with a local solution because the global Q-function is decomposed into local Q-functions that should be learned in a distributed manner. On the other hand, the present study uses a single reward that is derived from the total cumulative delay in an entire transportation network. A single agent finds an optimal traffic signal policy so that the total system delay can be minimized. At the end of every learning interval, the reward is set as +1 if the delay cumulated during the current interval is less than that cumulated during the previous interval and is set as -1 otherwise.

## 4. The structure of DGQN

4.1 An overview of graph convolution

A 2-D convolution cannot efficiently extract correlations between nodes in a graph. Kipf and Welling (2017) simplified a graph convolution method by confining the filters' operation to only the first-order neighbors. Repeating the graph convolution, however, extends to remote nodes from a target node. The working principle of the graph convolution is as follows.

Given a graph, $G = (V, E)$, a GCN requires two matrices as input. A feature matrix ($X \in \mathbb{R}^{N \times F^0}$) and an adjacency matrix ($A \in \mathbb{R}^{N \times N}$) is fed into a GCN. $V$ is a set of vertices (=nodes), $E$ is a set of edges, $N$ is the number of vertices, and $F^0$ is the number of given features for a node. A hidden layer of the GCN is represented as $H^l$=f($H^{l-1}, A$) where the initial hidden layer ($H^0$) is set to be $X$ and f is a propagation rule. How to choose the propagation rule is the key in the concept of GCN. Eq. (1) is the simplest form of a layer-wise propagation rule.

$$H^l = \sigma(AH^{l-1}Q^{l-1}) \quad (1)$$

In Eq. (1), $\sigma$ represents an activation function such as a sigmoid and a rectified linear unit (ReLU), and $Q^{l-1} \in \mathbb{R}^{F^{l-1} \times F^l}$ is a weight parameter matrix.

However, there are two typical problems regarding Eq. (1). First, multiplying an adjacency matrix with a hidden-feature matrix cannot transfer a node's own features to the next layer. Only features of its neighbor nodes can be passed to the next layer. To circumvent this drawback, an identity matrix is added to the given adjacency matrix ($\tilde{A} = I + A$). Second, feature values scale up as the layer-by-layer propagation repeats. The average normalization of the adjacency matrix ($\tilde{A}$) can be an option to sidestep this problem. A diagonal node-degree matrix ($D$) is prepared such that its diagonal element sums up the row values of the adjacency matrix ($D_{ii} = \sum_j \tilde{A}_{ij}$). The adjacency matrix is then normalized by multiplying the inverse of the diagonal matrix and the given adjacency matrix ($D^{-1}\tilde{A}$), and thus all rows of the resultant matrix sum to one. On the other hand, instead of this average normalization, Kipf and Welling (2017) adopted a spectral normalization to secure a greater dynamic for the propagation rule ($D^{-\frac{1}{2}}\tilde{A}D^{-\frac{1}{2}}$). The spectral normalization scheme divided an element of the adjacency matrix by the row sum and the column sum $\frac{\tilde{A}_{ij}}{\sqrt{D_{ii}}\sqrt{D_{jj}}}$), whereas the average normalization scheme divided an element only by the row sum in ($\frac{\tilde{A}_{ij}}{D_{ii}}$). The final propagation rule established by Kipf and Welling (2017) takes the form of Eq. (2).

$$H^{l} = \sigma e\left(D^{-\frac{1}{2}}\tilde{A}D^{-\frac{1}{2}}H^{l-1}Q^{l-1}\right) \quad (2)$$

The concept of this graph convolution, however, is not appropriate for a case where the intensity of connection varies by adjacent edges. The propagation of traffic states is a typical example wherein the influence of a road segment's traffic state is exerted differently on its first-order neighbors. A fixed adjacency matrix cannot accommodate such traffic propagation. Recently, some researchers utilized graph-attention networks to assign different weights to neighboring nodes and to learn the weights from data (Veličković et al., 2017; Zhang et al., 2018; and, Abu-El-Haija et al., 2018). Their model, however, is redundant when applied to a fixed-graph topology since it should train parameters for all possible connections between nodes. This type of a full-attention mechanism is inefficient for a problem involving a fixed graph such as a transportation network.

In our previous study, Yu et al. (2020), we devised a novel graph convolution scheme for a constant transportation network to overcome the redundancy and forecasted traffic speeds on an area-wide scale. This scheme was adapted to compose a DGQN for RL-based traffic signal control. The incorporation of a network topology into a Q-function facilitates the approximation of true action values from traffic states. The present study recomposed a transportation network such that vertices and edges represented lane groups and the connections between them, respectively. An adjacency matrix ($A^{l}$) was prepared such that the row and column dimension of the adjacency matrix was set equal to the total number of lane groups in the testbed. For the row of a specific lane group, a non-zero attention value was assigned only for columns corresponding to first-order downstream and upstream neighbors (see Fig. 1). By doing so, different traffic propagation patterns were accommodated. That is, the traffic state of a lane group can be differently affected by upstream or downstream lane groups.

$$A_{ij}^{l} = \text{softmax}(\theta_{ij}^{l}) = \frac{e^{\theta_{ij}^{l}}}{\sum_{k\in\mathfrak{N}_i} e^{\theta_{ik}^{l}}} \quad (1)$$

$A_{ij}^{l}$ is an element of an adjacency matrix ($A^{l}$) for the $l^{th}$ hidden layer, $\theta_{ij}^{l}$ is a weight parameter to be learned, and $\mathfrak{N}_i$ is a set of lane groups that are directly connected to the lane group $i$.

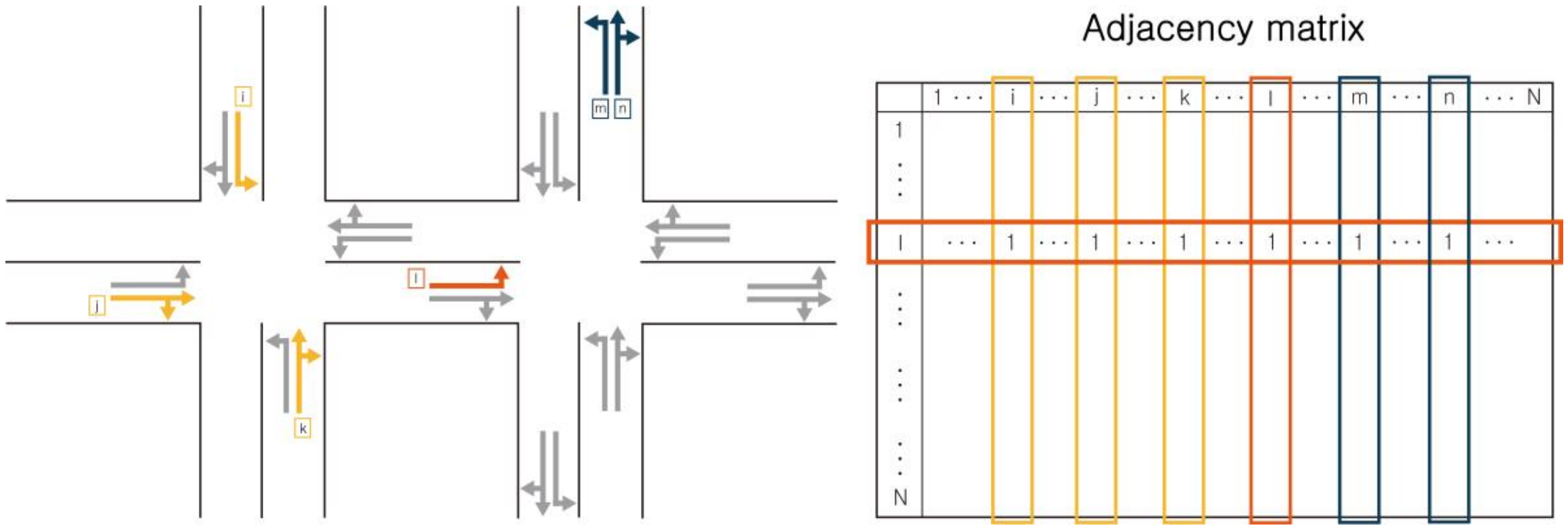

Fig. 1. The concept of adjacency matrix. The fixed adjacency matrix (right) is used as a mask to constitute the parameterized adjacency matrices in a DGQN

Repeating a graph convolution makes it possible to recognize the influence of lane groups in remote intersections. For each graph convolution in a DGQN, a different adjacency matrix was used with the position of non-zero cells maintained, so that the influence of neighboring lane groups could be differentiated by spatio-temporal domain. The proposed graph convolution is different from the original one in that connection intensities within an adjacency matrix are not fixed but parameterized.

## 4.2 The architecture of a DGQN

### 4.2.1 The former portion of a DGQN to elicit features from states

The former portion of the proposed DGQN was designed to extract features from input states. The portion was built on novel graph convolutions with multiple adjacency matrices to reflect the different spatio-temporal effects of neighboring lane groups by connection order. As far as we could ascertain, the present study is the first traffic signal control attempt to use a Q-network based on graph convolutions with parameterized adjacency matrices.

Fig. 2 (a) depicts the recursive expression of graph convolutions that constitutes a former portion of the DGQN. The row dimension for input ($N$) was set equal to the number of lane groups in the testbed, and the column dimension ($P$) of the input was set equal to the number of state variables. The number of graph convolutions was varied to consider the different spatio-temporal influences of input states. Four-dimensional weight parameters ($\theta_{ij}^{kt}$) were included to differentiate the impact of input states for different time intervals. A tensor ($\boldsymbol{S}_t = [s_{t-2}\ s_{t-1}\ s_t]$) that concatenated the vectors of three previous time intervals was chosen as the RL state at time $t$, and each input vector was separately fed to subsequent graph convolutions so that an older traffic state could be processed through a greater number of graph convolutions. Since the proposed model did not include separate weight matrices ($Q^l$), a dimension for hidden layers was maintained through all graph convolutions.

$$H_{t-2} = \sigma\left(A_\theta^{32}\sigma\left(A_\theta^{22}\sigma\left(A_\theta^{12}s_{t-2}\right)\right)\right) \tag{2-1}$$

$$H_{t-1} = \sigma\left(A_\theta^{21}\sigma\left(A_\theta^{11}s_{t-1}\right)\right) \tag{2-2}$$

$$H_t \ = \sigma\left(A_\theta^{10}s_t\ \right) \tag{2-3}$$

$$\boldsymbol{H}_t = [H_{t-2}\ H_{t-1}\ H_t\ ] \ \text{ for } \ t = 2, \ldots, T-2 \tag{2-4}$$

Eqs. (2-1) to (2-4) represent graph convolution layers to accommodate the spatio-temporal influence of input states on action values, wherein $\sigma$ is a softmax function, $s_t$ is an input state vector in $t$, $A_\theta^{kl}$ including $\{\theta_{ij}^{kl}\}$ is a parameterized adjacency matrix to accommodate the $k^{th}$ graph convolution for $s_{t-l}$, $H_t$ is a hidden layer from $s_t$ after going through graph convolutions, and $\boldsymbol{H}_t$ is a tensor that concatenates $H_{t-2}$, $H_{t-1}$, and $H_t$ along the first axis. The last hidden tensor after graph convolutions goes through a 2-D convolution, and the output tensor is flattened and fully connected to a dense layer. The dense layer acts as an input to the latter portion of the DGQN. The network architecture is presented in Fig. 2.

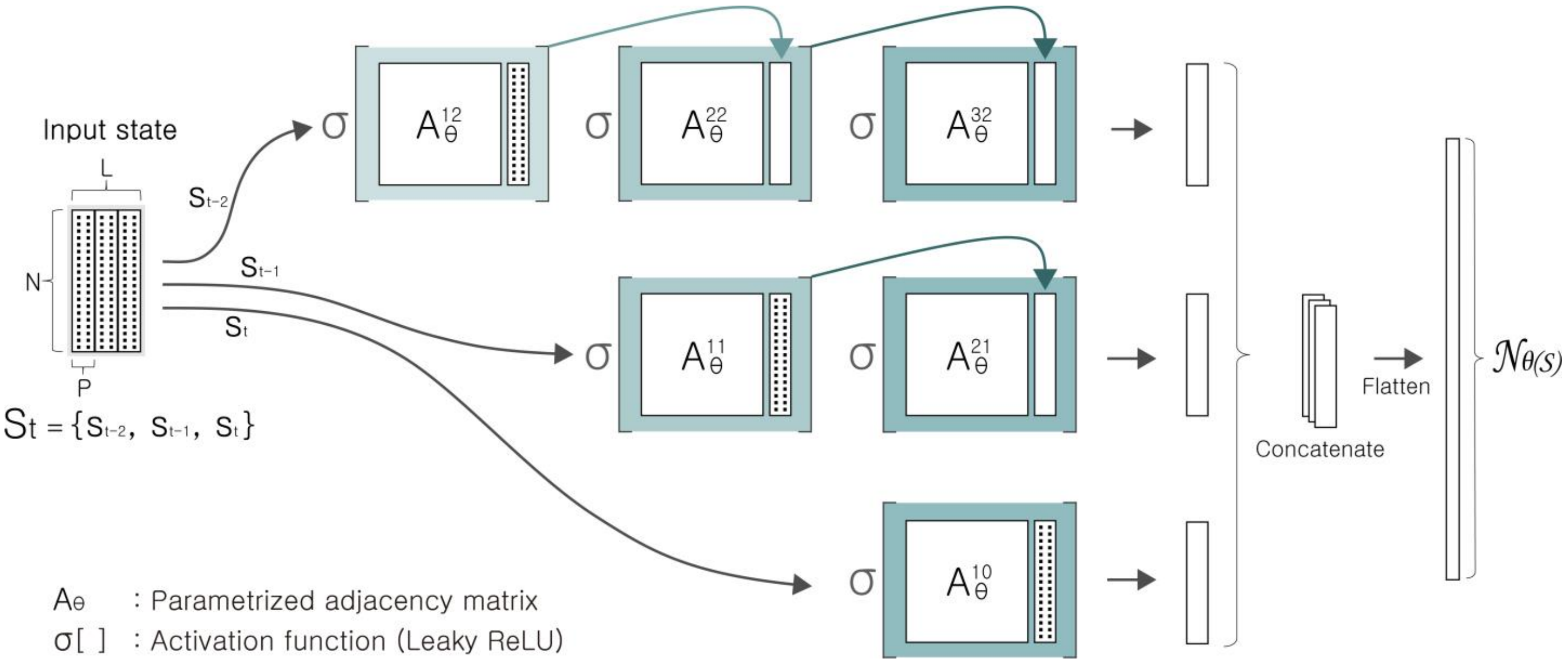

(a) The former portion for graph convolutions

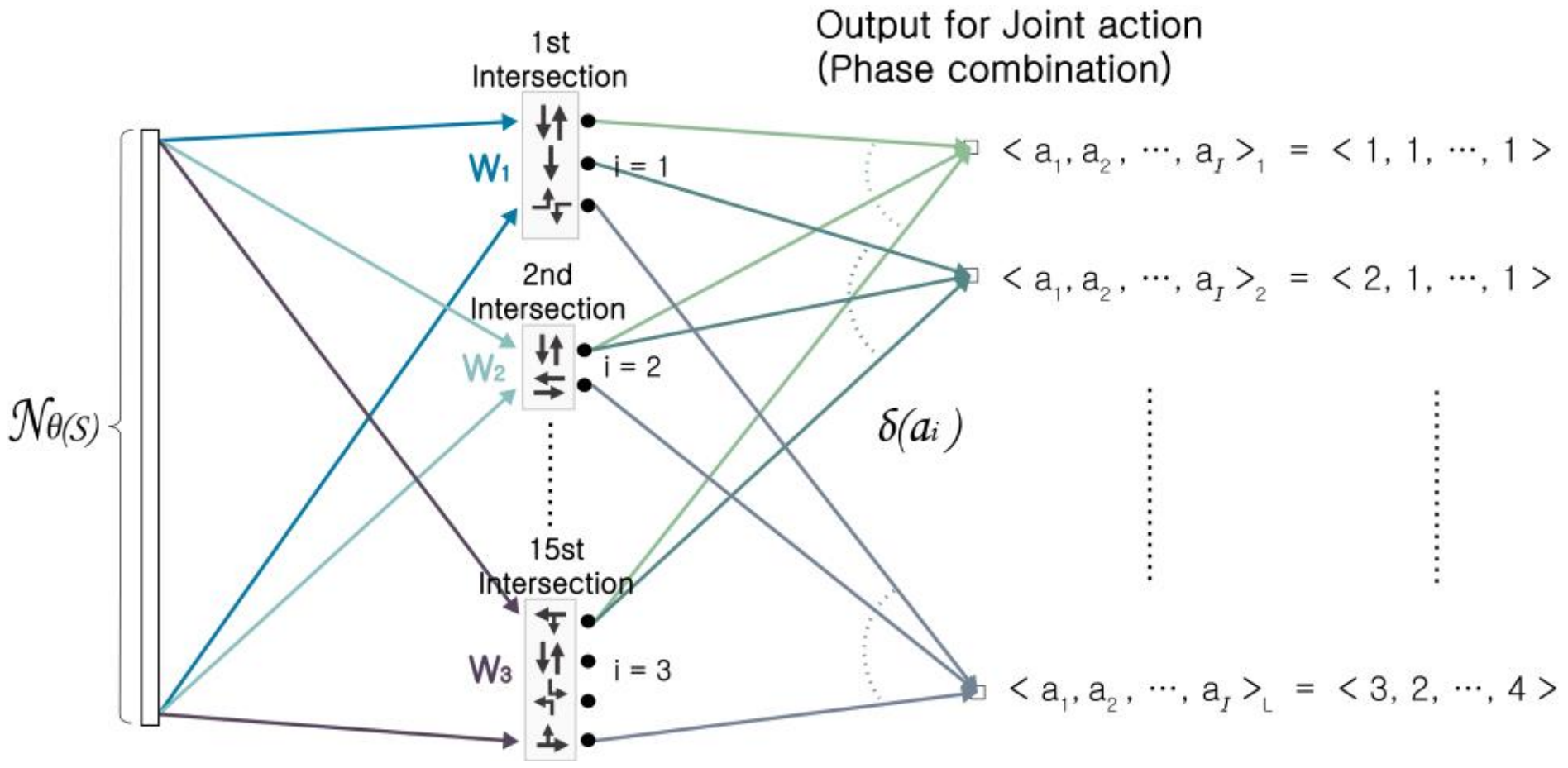

(b) The latter portion to accommodate the large action space

Fig. 2 The architecture of the proposed DGQN

4.2.2 The latter portion of the DGQN to accommodate a large action space

The latter portion of the proposed DGQN maps the abstracted features into actions. A conventional DQN takes a specific form whereby the number of output nodes should be the same as the size of the action space. Thus, if the number of joint actions increases, the specification cannot work properly. Our previous study overcame this problem by selectively fixing weight parameters between the last hidden layer and the output layer (Lee et al., 2019). This measure was mathematically expressed and embedded into the proposed DGQN. For brevity, the subscript for time will be dropped from all notations hereafter.

$$Q(\boldsymbol{S}, < a_1, \dots a_I > | \boldsymbol{\theta}, < \boldsymbol{W}_1, \dots, \boldsymbol{W}_I >) = \sum_{i=1}^{I}\sum_{j=1}^{\Phi} \boldsymbol{\mathcal{N}}_{\boldsymbol{\theta}}^{T}(\boldsymbol{S})\, \mathrm{w}_j^i \delta_j(a_i) = \sum_{i=1}^{I} \boldsymbol{\mathcal{N}}_{\boldsymbol{\theta}}^{T}(\boldsymbol{S}) \boldsymbol{W}_i \boldsymbol{\delta}(a_i) \qquad (3)$$

$I$ = number of intersections to be jointly controlled

$\Phi$ = max. number of feasible signal phases for an intersection

$N$ = number of lane groups (=77 for the present testbed)

$P$ = number of attributes to represent a state (=2 for the present study including delay and queue length)

$L$ = number of time lags (=3 for the present study)

$\boldsymbol{S}$= $[N \times P \times L]$ state tensor representing the traffic states of the entire transportation network during the present and $(L-1)$ previous intervals.

$< a_1, \dots a_I >$ = a joint signal phase (=action) with each $a_i \in \{1,..,\Phi\}$ representing a specific signal phase for the $i^{th}$ intersection

$M$ = number of nodes for the last FC layer

$\boldsymbol{\mathcal{N}_\theta}\,(\boldsymbol{S})$ = $[M \times 1]$ tensor corresponding to the last FC layer of the former portion of the DGQN

$\boldsymbol{\mathcal{N}_\theta^T}(\boldsymbol{S})$ = the transposed $\boldsymbol{\mathcal{N}_\theta}\,(\boldsymbol{S})$

$\boldsymbol{W}_i$ = $\{\boldsymbol{w}_{kj}^i\}$, $[M \times \Phi]$ weight matrix that includes connections between the last FC layer and the de-facto output nodes for the $i^{th}$ intersection, $k = 1, \dots, M$ , j= $1, \dots, \Phi$ and $i = 1, \dots, I$

$\mathrm{w}_j^i$ = $\left[w_{1j}^i, \dots, w_{Mj}^i\right]^T$, $[M \times 1]$ vector denoting the $j^{th}$ column of the matrix $\boldsymbol{W}_i$, $j = 1, \dots, \Phi$ and $i = 1, \dots, I$

$\boldsymbol{\delta}(a_i)$ = $\left[\delta_1(a_i), \dots, \delta_\Phi(a_i)\right]^T$, $[\Phi \times 1]$ indicator of the vector used to choose a phase given to the $i^{th}$

$\delta_j(a_i) = \begin{cases} 1 \;\; If\ a_i = the\ j^{th}\ signal\ phase \text{ of the intersection i} \\ 0 \qquad\qquad\qquad Otherwise \end{cases}$ , j $= 1, \dots, \Phi$ and $i = 1, \dots, I$

In Eq. (3), $\boldsymbol{\mathcal{N}_\theta}\,(\boldsymbol{S})$ is last hidden layer of the former portion of the DGQN following the graph convolutions. This layer can be interpreted as a function of an input traffic state ($\boldsymbol{S}$), given that $\boldsymbol{\theta}$ is a set of parameters $\{\theta_{ij}^{kl}\}$ associated with former graph convolutions to extract features from the traffic state. Elements of the indicator vector $[\boldsymbol{\delta}(a_i)]$ should meet the following conditions to guarantee that only a single signal phase is assigned to each intersection.

$$\sum_{j=1}^{\Phi} \delta_j(a_i) = 1,\ i = 1, \dots, I \tag{4}$$

The loss function represents the discrepancy between the target and the incumbent Q-networks in terms of the Bellman optimality equation. In Eq. (5), $\boldsymbol{\theta}^-$ and $< \boldsymbol{W_1^-}, \dots, \boldsymbol{W_1^-} >$ are parameters of the target Q-network that are fixed when updating the parameters of the incumbent Q-function. To get the target parameters, the parameters of the incumbent Q-function are periodically copied on a long-term basis while learning. A replay set was maintained so that prior experiences $\{(\boldsymbol{S}, < a_1, \dots a_I >, \boldsymbol{S}', r)\}$ could be stored. The loss function was minimized for each batch of examples taken randomly from the replay set. The loss function can be reduced to Eq. (6) by substituting Eq. (3) for Q-functions and introducing a batch scheme instead of general expectations. In Eq. (6), $B$ represents the batch size, and $\left(\boldsymbol{S}_b, < a_{1b}, \dots a_{Ib} >, \boldsymbol{S}_b^{'}, r_b\right)$ is the $b^{th}$ example in the batch. It is a great advantage that the maximum of the target Q-function can be obtained only by adding the largest node values in the de-facto output layer across all intersections without full enumeration of joint actions.

$$\mathcal{L}(\boldsymbol{\theta}, < \boldsymbol{W}_1, \dots, \boldsymbol{W}_I >) = E_{(\boldsymbol{S}, <a_1,\dots a_I>, \boldsymbol{S}', r)}\left[\left(\underbrace{r + \gamma \max_{<a_1',\dots a_I'>} Q(\boldsymbol{S}', < a_1', \dots a_I' > |\boldsymbol{\theta}^-, < \boldsymbol{W}_1^-, \dots, \boldsymbol{W}_I^- >)}_{A\ constant\ value\ with\ the\ target\ Q-network} - Q(\boldsymbol{S}, < a_1, \dots a_I > |\boldsymbol{\theta}, < \boldsymbol{W}_1, \dots, \boldsymbol{W}_I >)\right)^2\right] \quad (5)$$

$$\approx \sum_{b=1}^{B}\left[\left(r_b + \gamma \max_{<a_1',\dots a_I'>} \sum_{i=1}^{I} \boldsymbol{\mathcal{N}}_{\boldsymbol{\theta}^-}^T(\boldsymbol{S}_b')\boldsymbol{W}_i^-\boldsymbol{\delta}(a_i') - \sum_{i=1}^{I} \boldsymbol{\mathcal{N}}_{\boldsymbol{\theta}}^T(\boldsymbol{S}_b)\boldsymbol{W}_i\boldsymbol{\delta}(a_{ib})\right)^2\right]/B \quad (6)$$

Fig. 2 depicts the architecture of the proposed DGQN. The same architecture was applied to the target Q-network. For convenience, the former and later portions of the network are separately drawn in Fig. 2. The first hidden layer of Fig. 2 (b) is the same as the last FC layer of Fig.2 (a).

## 5. Asynchronous algorithm to train DGQN

The architecture of a DGQN takes the form of a single-agent RL method but can accommodate a large state and action spaces without the need of an exorbitant amount of computer memory. However, this does not mean that the proposed DGQN can be trained using only limited examples of state-action pairs. Basically, for convergence a DGQN must go through as many transition experiences as possible while training, which requires a considerable amount of time. To reduce the computation time in the present study, multiple actor-learners were mobilized in parallel, and a RL environment was copied to them so that a separate traffic simulation could be carried out for each actor-learner. At a given simulation time, each actor-learner had a traffic condition that differed from that of their counterparts, which made it possible to avoid correlations between consecutive traffic states when a single traffic simulation would be used.

A common Q-network was set up and asynchronously updated by multiple actor-learners (see Fig. 3), each of which learned from its own environment. A target Q-network was also shared by multiple actor-learners. Each actor-learner managed its own replay set unlike the original A3C algorithm wherein no replay set is used. This scheme required a greater amount of computer memory but was advantageous in facilitating the convergence. An actor-learner in an asynchronous setting should not be confused with an agent in a multi-agent RL algorithm. Basically, actor-learners must be independent of each other in that an agent's action only affects its own environment, whereas agents in a multi-agent RL algorithm must be dependent in that an agent's action changes a single shared environment. An outline of the proposed asynchronous algorithm for each actor-learner thread is shown in Table 1.

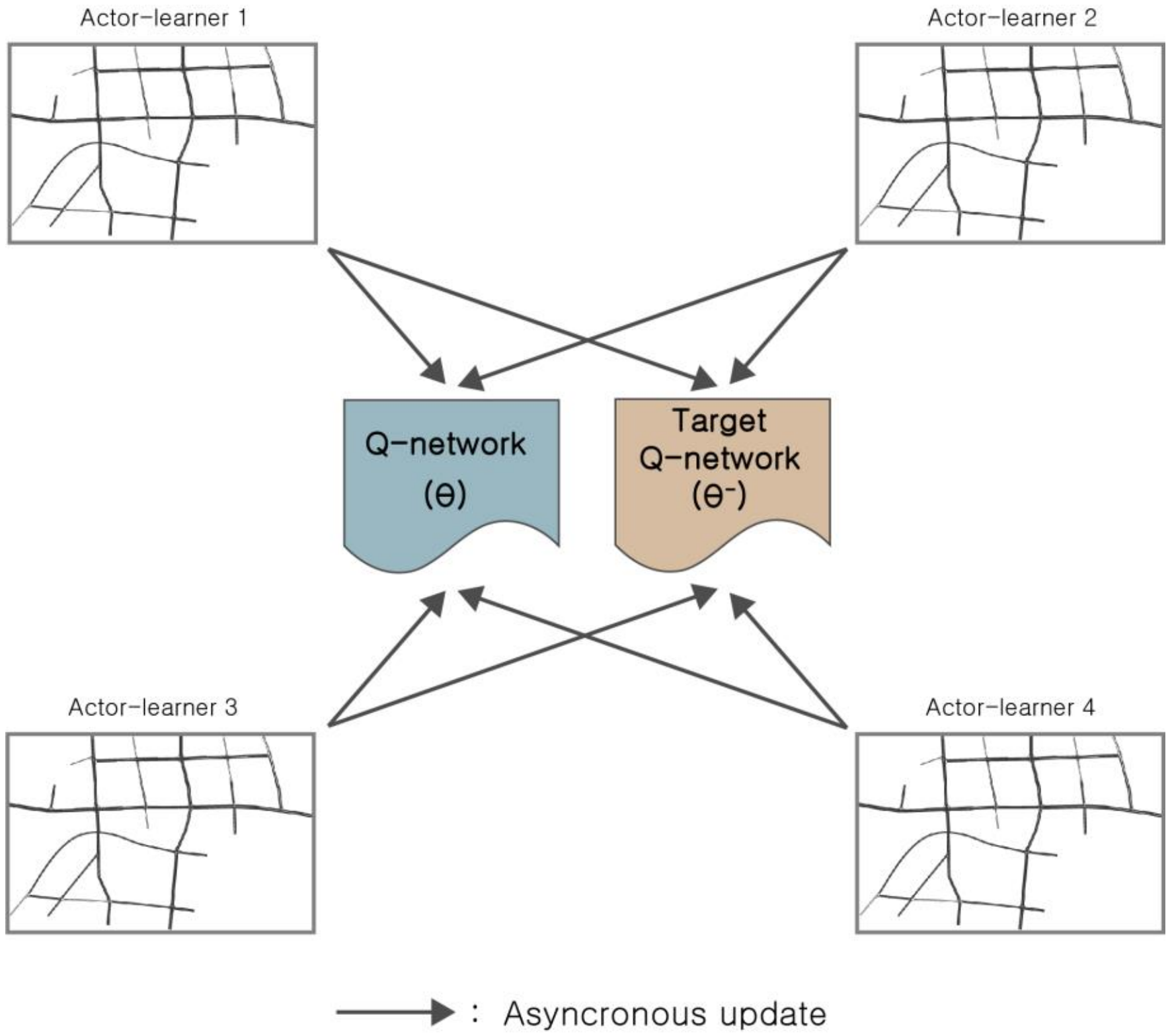

Fig. 3. The concept of asynchronous updating of a shared Q-network

The simulation experiment was performed under the following computing conditions. The main memory was 256 GB. Python main-logic and traffic-simulation software (Vissim v10.0) was implemented on two CPUs with the following specifications: AMD Ryzen Threadripper 3960X Processor @ 3.8GHz. There were 24 available CPU cores, which fell under the maximum number of cores (=32) that the traffic simulator allows. The proposed DGQN was trained on a single GPU, which was a NVIDIA GeForce GTX 1080Ti with 11 GB of GDDR5 memory. A graphic accelerator was used to train the DGQN, whereas the animation was run on CPU cores. It took an average of 4 minutes to run a single episode. The total computation time for 2,000 episodes was tantamount to 40 hours.

Table 1. Asynchronous algorithm for each actor-learner thread to train the proposed Q-network

//Assume global shared parameters $(\boldsymbol{\theta}^{-}, <\boldsymbol{W}_1^{-}, \ldots, \boldsymbol{W}_I^{-}>)$ and $(\boldsymbol{\theta}, <\boldsymbol{W}_1, \ldots, \boldsymbol{W}_I>)$

Initialize replay set

Set counter $C = 0$

Implement the warm-up simulation for $T_{Initial}$

**repeat**

Set simulation time $T = T_{Initial}$.
Initialize traffic simulator
Choose initial state $\boldsymbol{S}$
**while** $T < T_{max}$ and traffic simulation is not over[1]

$\epsilon = (\varepsilon_{max} - \varepsilon_{min})\mathrm{e}^{-(C/E_i)^2} + \varepsilon_{min}$ (7)

[2]Choose a joint action $< a_1, \ldots a_I >$ with $\epsilon$-greedy policy using

$$Q(\boldsymbol{S}, < a_1, \ldots a_I > | \boldsymbol{\theta}, < \boldsymbol{W}_1, \ldots, \boldsymbol{W}_I >) = \sum_{i=1}^{I} \boldsymbol{\mathcal{N}}_{\boldsymbol{\theta}}^{T}(\boldsymbol{S})\boldsymbol{W}_i\boldsymbol{\delta}(a_i).$$

Proceed traffic simulation with the chosen joint phases during $(\Delta t - T_a)$

Receive new state $\boldsymbol{S}'$ and reward $r$

Store $\left(\boldsymbol{S}, < a_1, \ldots a_I >, \boldsymbol{S}', r\right)$ in replay set

**If** the size of the replay set reaches $\boldsymbol{D}$ **then** remove the samples in a FIFO fashion.

Choose $B$ samples $\left(\boldsymbol{S}_b, < a_{1b}, \ldots a_{Ib} >, \boldsymbol{S}'_b, r_b\right)$ from the replay set

$$y_b = r_b + \gamma \max_{<a'_1, \ldots a'_I>} \sum_{i=1}^{I} \boldsymbol{\mathcal{N}}_{\boldsymbol{\theta}^{-}}^{T}\left(\boldsymbol{S}'_b\right)\boldsymbol{W}_i^{-}\boldsymbol{\delta}\left(a'_{ib}\right) \; for \; \mathrm{b} = 1, \ldots, B$$

Compute gradients: $$d\boldsymbol{\theta} = \frac{\partial \sum_{b=1}^{B}\left(y_b - \sum_{i=1}^{I} \boldsymbol{\mathcal{N}}_{\boldsymbol{\theta}}^{T}(\boldsymbol{S}_b)\boldsymbol{W}_i\boldsymbol{\delta}(a_{ib})\right)^2}{\partial \boldsymbol{\theta}}$$

$$d\boldsymbol{W}_{\mathrm{i}} = \frac{\partial \sum_{b=1}^{B}\left(y_b - \sum_{i=1}^{I} \boldsymbol{\mathcal{N}}_{\boldsymbol{\theta}}^{T}(\boldsymbol{S}_b)\boldsymbol{W}_i\boldsymbol{\delta}(a_{ib})\right)^2}{\partial \boldsymbol{W}_{\mathrm{i}}} \; for \; i = 1, \ldots, I$$

Perform asynchronous update $\boldsymbol{\theta}$, $< \boldsymbol{W}_1, \ldots, \boldsymbol{W}_I >$ using $d\boldsymbol{\theta}$ and $< d\boldsymbol{W}_1, \ldots, d\boldsymbol{W}_I >$
$\boldsymbol{S} = \boldsymbol{S}'$

$T = T + (\Delta t - T_a)$
$C = \mathrm{C} + 1$
Implement amber phase during $T_a$ for lane groups with a signal change.
Implement the current phase during $T_a$ for lane groups with a signal that does not change.
$T = T + T_a$

**If** $C$ mod $I_{target} == 0$ **then**

Update the target Q-network $\boldsymbol{\theta}^{-} = \boldsymbol{\theta}$, $< \boldsymbol{W}_{1\mathrm{e}}^{-}, \ldots, \boldsymbol{W}_{I}^{-} > = < \boldsymbol{W}_{1}, \ldots, \boldsymbol{W}_{I} >$

**until** $\epsilon \approx \varepsilon_{min}$

[1.] Traffic simulation is terminated when either an entry approach is full of vehicles or the total delay in a transport network during a single period exceeds the threshold (=16,000 seconds)

[2.] Finding the maximum is trivial based on the following relationship

$$\max_{<a_1,\ldots a_I>} \sum_{i=1}^{I} \boldsymbol{\mathcal{N}}_{\boldsymbol{\theta}}^{T}(\boldsymbol{S})\boldsymbol{W}_i\boldsymbol{\delta}(a_i) = \sum_{i=1}^{I} \max_{a_i \in \{1,..,\Phi\}} \boldsymbol{\mathcal{N}}_{\boldsymbol{\theta}}^{T}(\boldsymbol{S})\boldsymbol{W}_i\boldsymbol{\delta}(a_i).$$

## 6. Simulation experiments

6.1 A description of the testbed

A handicap of RL-based traffic control is that the algorithm cannot be trained in the real world. A robust simulator is necessary to train the algorithm. Vissim v10.0, a commercial traffic simulator, was chosen to mimic real-world traffic conditions. On the other hand, the testbed was chosen among real transportation networks. The testbed lies in the southwest area of Seoul, Korea as shown in Fig. 4. The network includes 15 intersections to be jointly controlled in the present study. Intersections that are not numbered indicate an unsignalized intersection or a grade-separated intersection without traffic lights. Fig. 4 also lists the current signal phases for each intersection in the morning peak hours. The reason some 4-leg intersections (①, ②, ⑨, ⑬, and ⑭) have only two or three phases is because they have a major road that either over- or under-passes a minor road. The cycle length ranges from 120 to 190 seconds.

The current phases were used as feasible phases for the present RL algorithm. That is, the currently available phases were implemented without a constant order for every time interval of the RL algorithm. The proposed RL-based traffic signal controller does not depend on the cycle length. The order of the signal phases varied according to traffic conditions. Thus, the controller must provide drivers with a relevant warning, so that they will be cognizant of variations in the signal order.

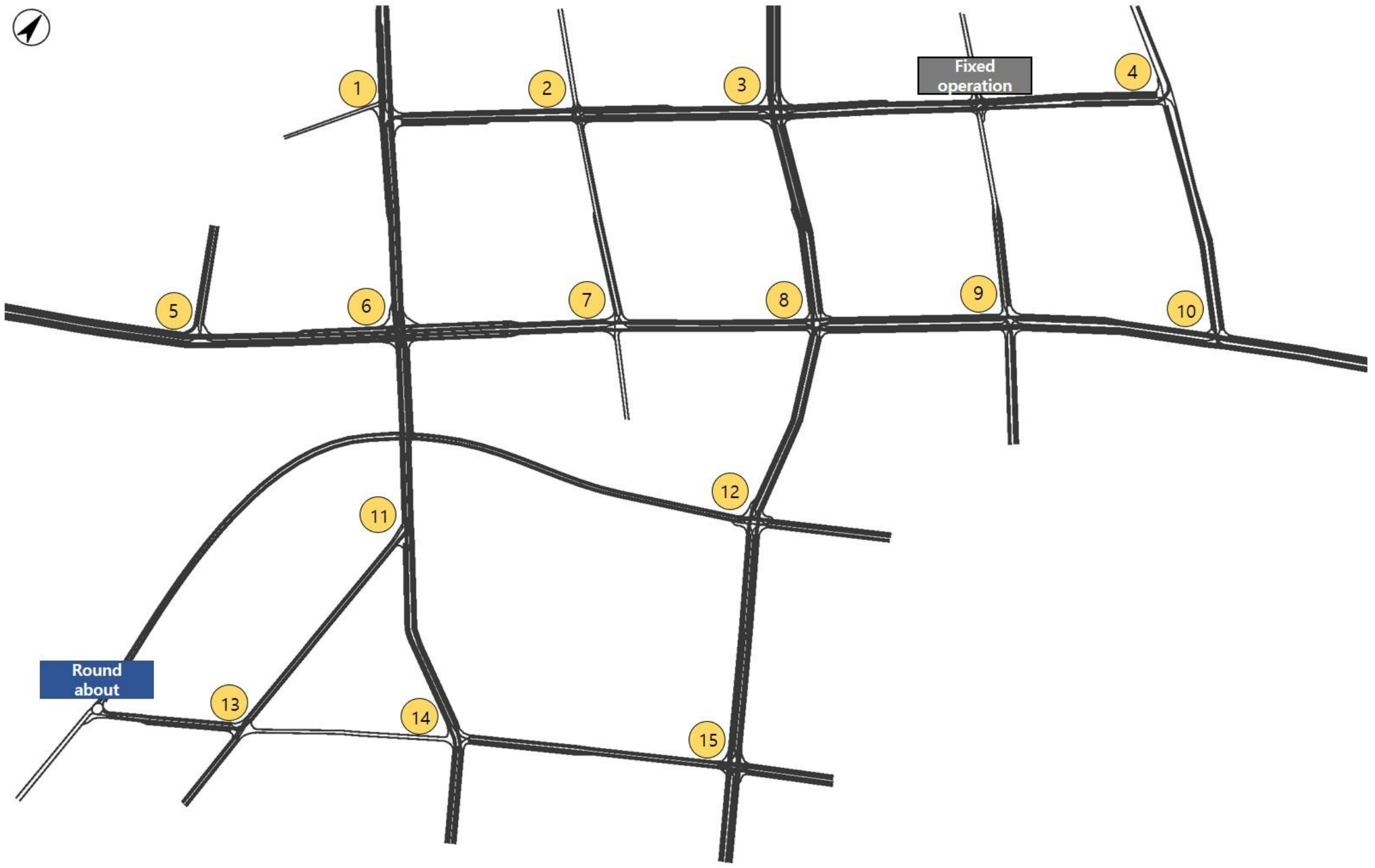

| | Inter section 1 | Inter section 2 | Inter section 3 | Inter section 4 | Inter section 5 | Inter section 6 | Inter section 7 | Inter section 8 | Inter section 9 | Inter section 10 | Inter section 11 | Inter section 12 | Inter section 13 | Inter section 14 | Inter section 15 |
|---|---|---|---|---|---|---|---|---|---|---|---|---|---|---|---|
| Phase1 | 82 | 40 | 45 | 55 | 120 | 37 | 93 | 82 | 109 | 107 | 100 | 45 | 107 | 64 | 30 |
| Phase2 | 48 | 80 | 20 | 35 | 39 | 56 | 40 | 82 | 25 | 19 | 50 | 25 | 33 | 39 | 55 |
| Phase3 | 40 | | 37 | 30 | 21 | 51 | 37 | 82 | 26 | 44 | 40 | 40 | | 37 | 25 |
| Phase4 | | | 18 | | | 36 | | 82 | | | | 25 | | | 30 |
| Cycle(s) | 170 | 120 | 120 | 120 | 180 | 180 | 170 | 180 | 160 | 170 | 190 | 135 | 140 | 140 | 140 |

Fig. 4 Testbed for simulation experiments

## 6.2 Setting up traffic simulations

Traffic simulation requires the traffic volumes in the entry links of the testbed and the turning rates of traffic flows at the stop lines of intersections. The present experiment used input data that are compatible with real-world traffic conditions. The data were excerpted from the final report of a traffic impact analysis study that had been implemented in the testbed. The average entry traffic volumes and the average turning rates are shown in Tables 2 and 3, respectively.

Table 2. Average traffic volumes in the entry links to the testbed during the morning peak hours

| Intersection | Direction | Entry Volume (veh/h) |
|---|---|---|
| 1 | East-bound | 81 |
| 1 | South-bound | 2,506 |
| 2 | South-bound | 259 |
| 3 | South-bound | 169 |
| 4 | South-bound | 494 |
| 5 | East-bound | 2,511 |
| 5 | South-bound | 352 |
| 7 | North-bound | 13 |
| 9 | North-bound | 236 |
| 10 | West-bound | 3,107 |
| 12 | West-bound | 972 |
| 13 | North-bound | 485 |
| 14 | North-bound | 871 |
| 15 | West-bound | 231 |
| 15 | North-bound | 457 |
| Fixed operation | South-bound | 183 |
| Roundabout | North-bound | 575 |

Table 3. Average turning rates of traffic flows during the morning peak hours

| Intersection | East-bound | | | West-bound | | | North-bound | | | South-bound | | |
|---|---|---|---|---|---|---|---|---|---|---|---|---|
| | Left turn (%) | Through (%) | Right turn (%) | Left turn (%) | Through (%) | Right turn (%) | Left turn (%) | Through (%) | Right turn (%) | Left turn (%) | Through (%) | Right turn (%) |
| 1 | 83 | - | 17 | 24 | - | 76 | - | 85 | 15 | 14 | 58 | 28 |
| 2 | 4 | 80 | 16 | 9 | 86 | 5 | 26 | 29 | 45 | 59 | 23 | 18 |
| 3 | 11 | 62 | 27 | 20 | 73 | 7 | 65 | 10 | 25 | 33 | 35 | 32 |
| 4 | 57 | 53 | - | - | - | - | 46 | 55 | - | 44 | - | 56 |
| 5 | 13 | 87 | - | - | 82 | 18 | - | - | - | 49 | - | 51 |
| 6 | 20 | 6 | 74 | 55 | 19 | 26 | 43 | 46 | 11 | 11 | 83 | 6 |
| 7 | 11 | 81 | 8 | 93 | - | 7 | - | - | 100 | 35 | 14 | 51 |
| 8 | - | 81 | 19 | 3 | 69 | 28 | 37 | 23 | 40 | 19 | 71 | 10 |
| 9 | 6 | 87 | 7 | 7 | 87 | 6 | 68 | - | 32 | 7 | - | 93 |
| 10 | 7 | 93 | - | - | 83 | 17 | - | - | - | 67 | - | 33 |
| 11 | - | 94 | 6 | - | - | - | - | 100 | - | - | 53 | 47 |
| 12 | 37 | 45 | 18 | 26 | 32 | 42 | 10 | 59 | 31 | 25 | 38 | 37 |
| 13 | 71 | - | 29 | - | - | 100 | 16 | 70 | 14 | - | 59 | 41 |
| 14 | - | - | 100 | 46 | - | 54 | - | 82 | 18 | 15 | 82 | 3 |
| 15 | - | 31 | 69 | - | 74 | 26 | 11 | 73 | 16 | 2 | 81 | 17 |

The traffic volume of only the approaches entering the testbed was set as the input. Traffic volumes on the inner road segments were determined by the turning rates of the traffic flows at each stop line of intersection approaches. Each actor-learner shared a fixed set of average traffic flows at entry links and turning rates at each stop line of intersection approaches, which were observed in the morning peak hours. However, each actor-learner varied the traffic volumes within ±30% at the beginning of each episode to reflect the traffic conditions for different times of the day. In addition, the average turning rates of traffic volumes were periodically altered within ±30% during an episode in an effort to reflect the random fluctuations of real traffic volumes. Fig. 7 shows how traffic volumes and turning rates change for an episode. Because some entry approaches may be saturated due to randomly varied traffic volumes while implementing a RL algorithm, an episode for each actor-learner was terminated when at least one of the entry approaches was full of vehicles and could not receive a new vehicle from outside the transport network.

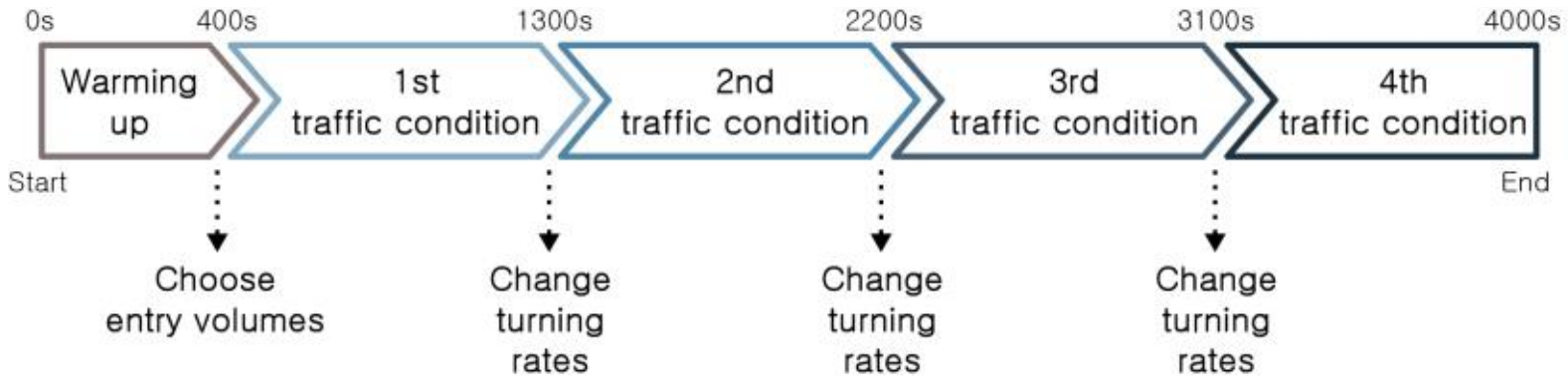

Fig. 7. Varying traffic conditions for an episode. Traffic simulation during the warm-up period was based on the average entry traffic volumes and on the average turning rates.

## 6.3 Setting up the hyper-parameters of an RL algorithm

Table 4 shows the hyper-parameters used to implement the present RL algorithm. The simulation time for each episode was set at 4,000 seconds, which included 400 seconds of idling periods for warm-up. The traffic simulation was implemented each second of the simulation clock, and a RL update was implemented every 20 seconds of the simulation clock. More concretely, a RL update was carried out after 17 seconds from the onset of every update period, and 3 seconds of time was provided for amber signals. The lane groups with signal phase changes at the current update received a 3-second amber phase on the simulation clock as the remaining lane groups continued their current phase.

Setting up the rates of exploration and exploitation is very important for the convergence of a RL algorithm. The exploration rate dwindled from the maximum (=1.0) down to the minimum (=0) while implementing a RL algorithm. The present study differentiated the exploration rate for each actor-learner. A different decaying parameter was applied to each actor-learner, as shown in Table 4. Empirical evidence showed that this scheme expedited the convergence. The target Q-network was updated for every 2,500 updates of the incumbent Q-network.

Table 4. Hyper-parameters for implementing an RL algorithm

| Hyper-parameter | Description | Applied value |
|---|---|---|
| $\Delta$ | Time step for traffic simulation | 1 second |
| $T_{max}$ | Simulation period for each episode | 4,000 seconds |
| $T_{initial}$ | Warming-up period for each episode | 400 seconds |
| $\Delta t$ | Time interval for RL algorithm | 20 seconds |
| $T_a$ | Amber time | 3 seconds |
| $\varepsilon_{max}$ | Initial probability of exploration | 1.0 |
| $\varepsilon_{min}$ | Final probability of exploration | 0 |
| $E_1$ | Decaying parameter of exploration probability for actor-learner1 | 2.3E6 |
| $E_2$ | Decaying parameter of exploration probability for actor-learner2 | 2.6E6 |
| $E_3$ | Decaying parameter of exploration probability for actor-learner3 | 2.9E6 |
| $E_4$ | Decaying parameter of exploration probability for actor-learner4 | 3.2E6 |
| D | Size of replay memory set | 30,000 |
| $D_{initial}$ | Initial replay memory | 3,000 |
| B | Size of mini-batch | 32 |
| $I_{target}$ | Cycle for updating target Q-function | 2,500 iterations |

## 7. Results from simulation experiments

The objectives of the present study were three-fold. The first objective was to confirm whether the proposed DGQN, with modifications to accommodate a large action space, could effectively manage joint traffic signal control on an area-wide scale. The objective was accomplished since a RL model with the remedy to accommodate a large action space succeeded in jointly controlling 15 intersections. The performance comparison with multi-agent RL models was omitted because based on a small synthesized network our previous study had already proved that a modified DQN outperformed all other types of multi-agent RL algorithms (Lee et al., 2019). The second objective was to prove the utility of adopting graph convolutions with variable adjacency matrices. To verify this objective, several reference models were set up for comparison (see the section 7.1). The last objective involved demonstrating the efficiency of asynchronous updates for a DGQN. The evidence for the utility of an asynchronous update was clarified, because we failed in leading to a convergence of the DGQN with a single environment. In theory, it would be possible for a single actor-learner to obtain a convergence with a single environment if a long computing time was allowed. However, a DGQN did not lead to a convergence without asynchronous updating, even though it was trained on the same number of episodes as the sum of each actor-learner's episodes.

### 7.1 Selecting reference models

The current fixed operation of traffic signals in the testbed was chosen as a baseline (see Fig. 4). A simple RL algorithm that involved a DQN composed only of fully connected layers was chosen as the second reference (DQN-FC). Another DQN with the original graph convolutions that used a constant adjacency matrix was established as the last reference model (DQN-OGCN). The performances of these three references were compared with that of the proposed DGQN. For fair comparisons, the proposed scheme to accommodate a large action dimension was applied to the architectures of both the DQN-FC and the DQN-OGCN. Moreover, each reference network was designed so that the number of weight parameters could match that of the proposed DGQN. In particular, the former architecture of the DQN-OGCN was the same as that of the proposed DGQN, with the exception of a parameterized adjacency matrix.

For reference models and a DGQN, 600 episodes were implemented to test models trained for about 2,000 episodes. For each testing episode, the same rule used in training times was also adopted to generate traffic conditions.

### 7.2 The convergence of models

An asynchronous update was applied to training all the reference models as well as a DGQN. Fig. 8 shows the convergence of RL models. During training, an average reward was computed for every episode and plotted as implementing successive episodes. For brevity, the convergence of the first actor-learner was depicted, since those of the remaining actor-learners were not much different. The black solid line represents the moving average of the previous 10 average rewards, which clarifies the trend of convergence. The average reward of the former two models was based on graph convolutions and converged at 0.3, whereas the average reward for the DQN-FC with FC layers led to a higher value (=0.5). This was because the traffic delays of the latter model fluctuated more than those of the two former models during training.

Fig. 9 expounds upon this phenomenon. The total delay fluctuated more within each episode for a DQN-FC. If the total delay increased at any iteration of an episode, a DQN-FC successively received a positive reward that allowed it to recover the previous damage. On the other hand, there was no drastic change in the total delay while training a DGQN. Thus, a DGQN showed a more consistent pattern of alternating positive and negative rewards, which decreased the average rewards during training. As a result, the variance in traffic delays was smaller for a DGQN than that for a DQN-FC.

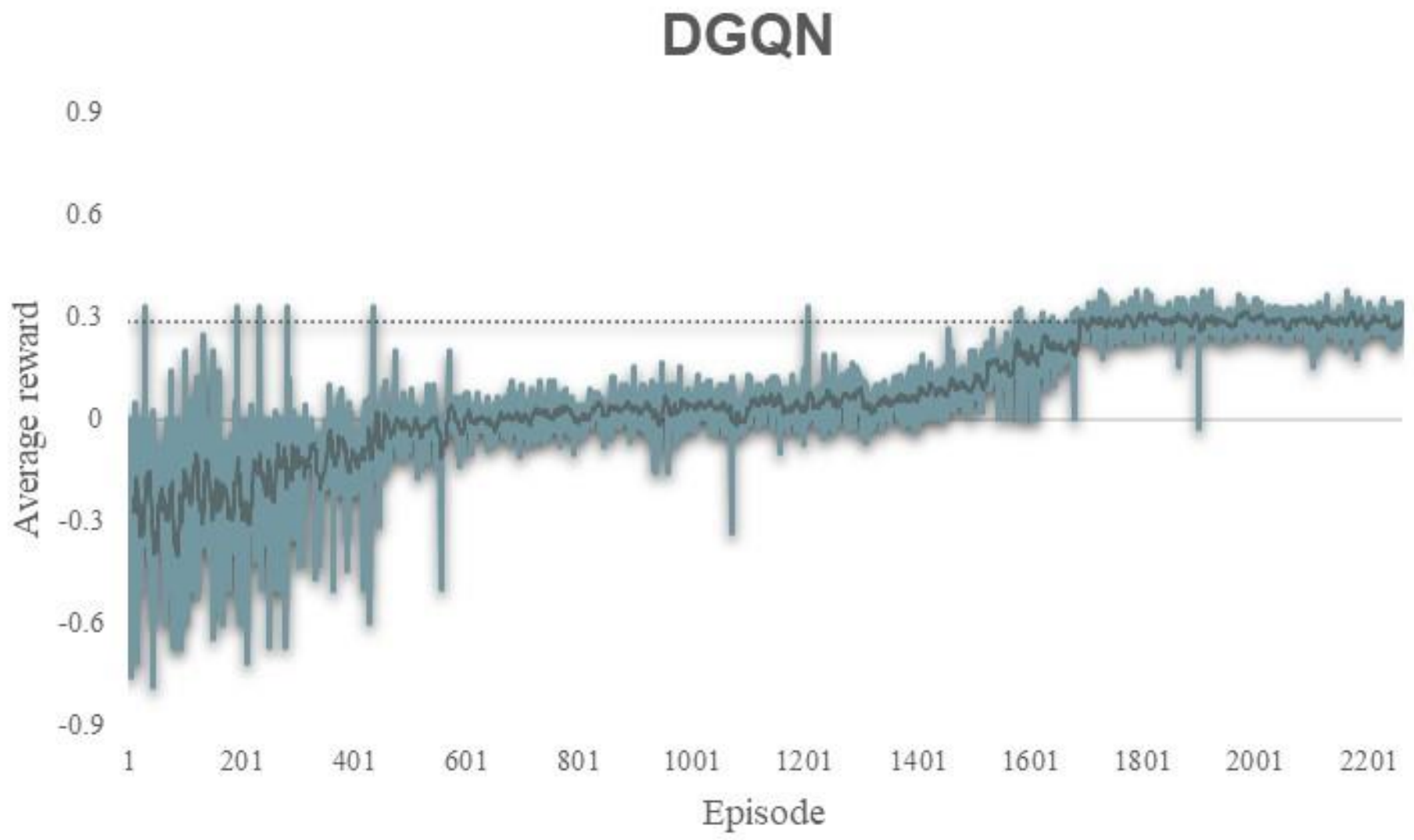

(a) Convergence in a DGQN

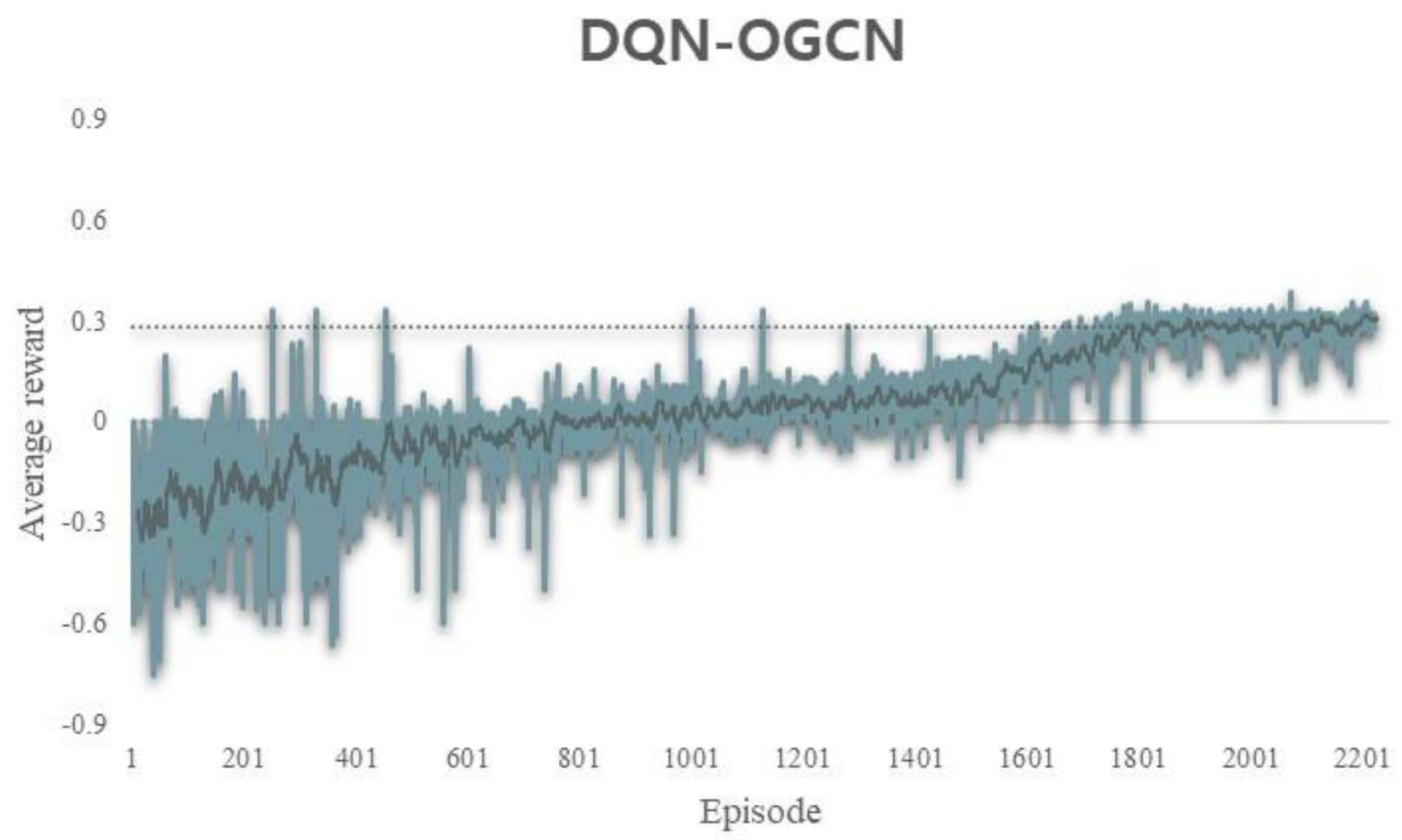

(b) Convergence in a DQN-OGCN

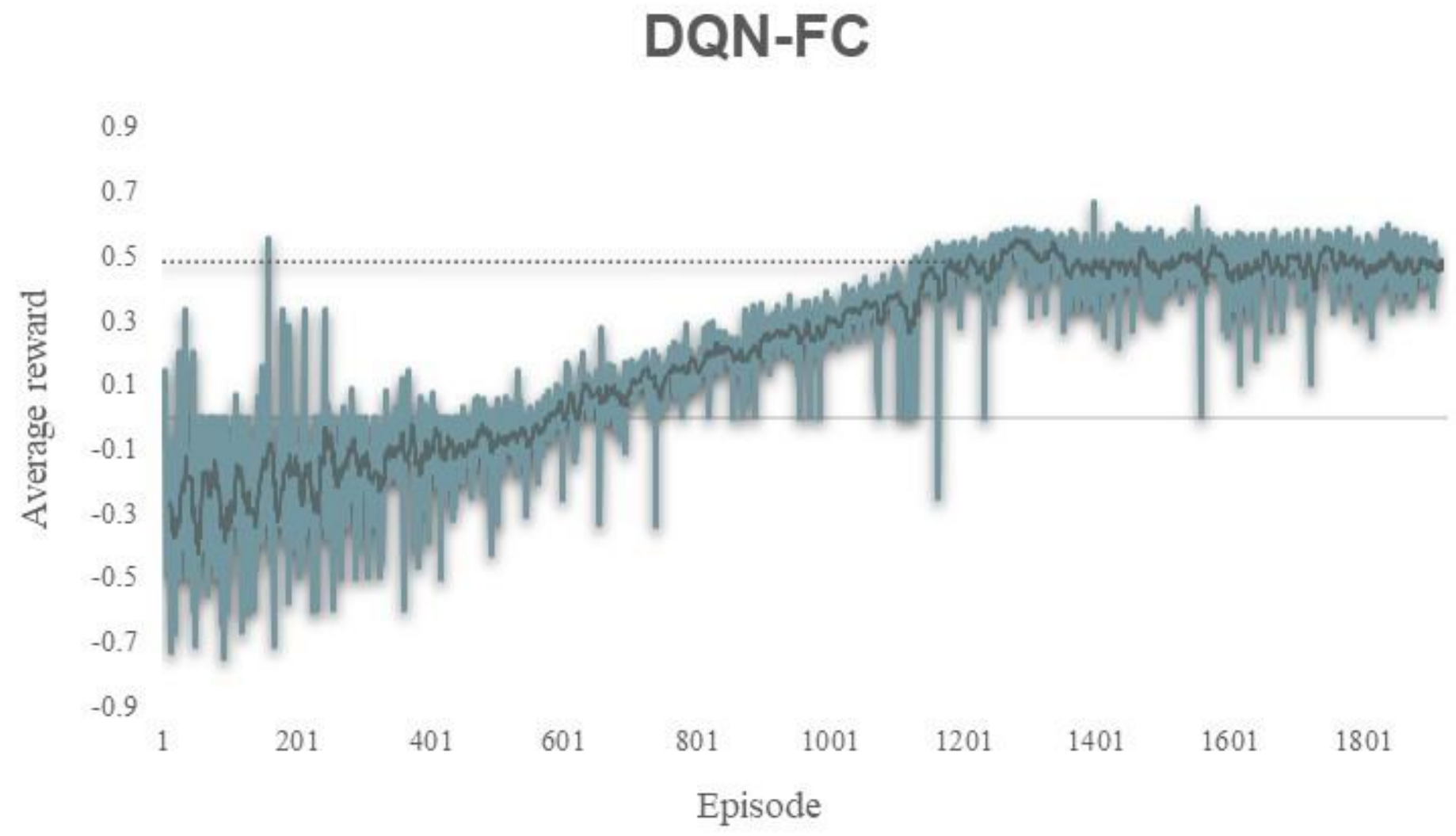

(c) Convergence in a DQN-FC

Fig. 8. Convergence in RL models

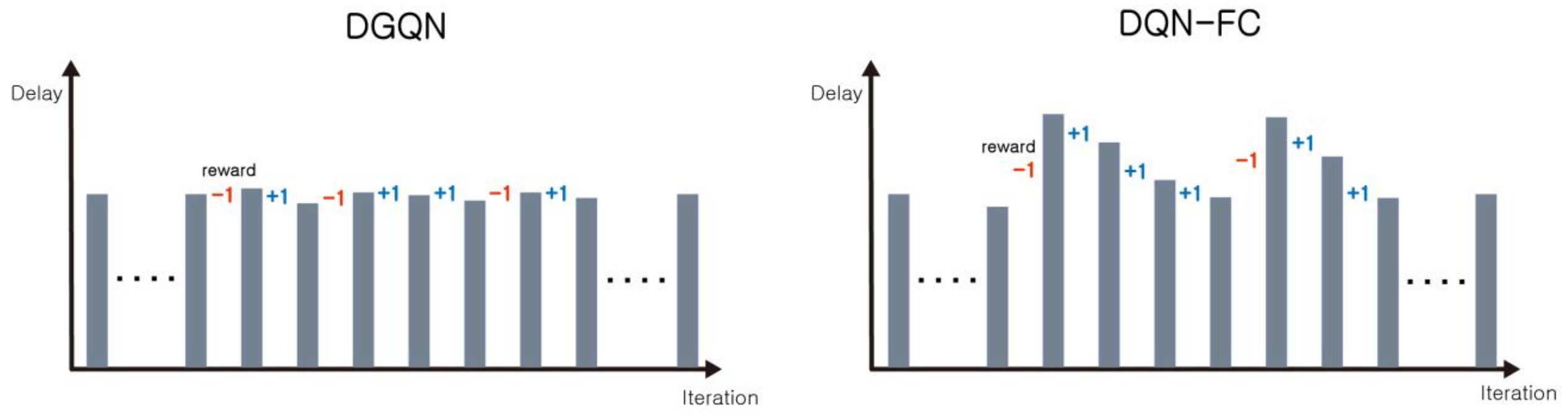

Fig. 9 The trend of rewards for different RL models

7.3 Performance comparisons

According to guidelines in the highway capacity manual, the total traffic delay and the maximum queue length were chosen to compare the performance of the proposed DGQN with those of reference models. For each model, 600 delay indices and 600 queue length indices were derived from testing episodes. Fig. 10 shows the probabilistic densities drawn from the 600 indices. When comparing models based on the mean of performance indices, the DGQN outperformed all other references. The DQN-OGCN ranked as the second-best model, which implies that graph convolutions were advantageous in recognizing spatio-temporal dependencies among traffic states in a transportation network. Regarding comparisons based on the variance of performance indices, the DGQN also proved to be more robust than the two other RL-based reference models, which means it resulted in the most consistent operation of traffic lights.

There was a distinct outcome with regard to the variance of the performance indices. A fixed operation had smaller variances than the two other RL-based reference models, even though it had larger average delays and queue lengths. This result is consistent with a finding by Casas (2017) that the output of Q-learning when applied to traffic signal control showed a larger variance. On the other hand, a DGQN had a smaller variance in traffic delays compared with an affixed operation. In this regard, the success of a DGQN in reducing the variance in delays translates to a great enhancement. Also, a DGQN had a similar variance in maximum queue lengths compared with a fixed operation.

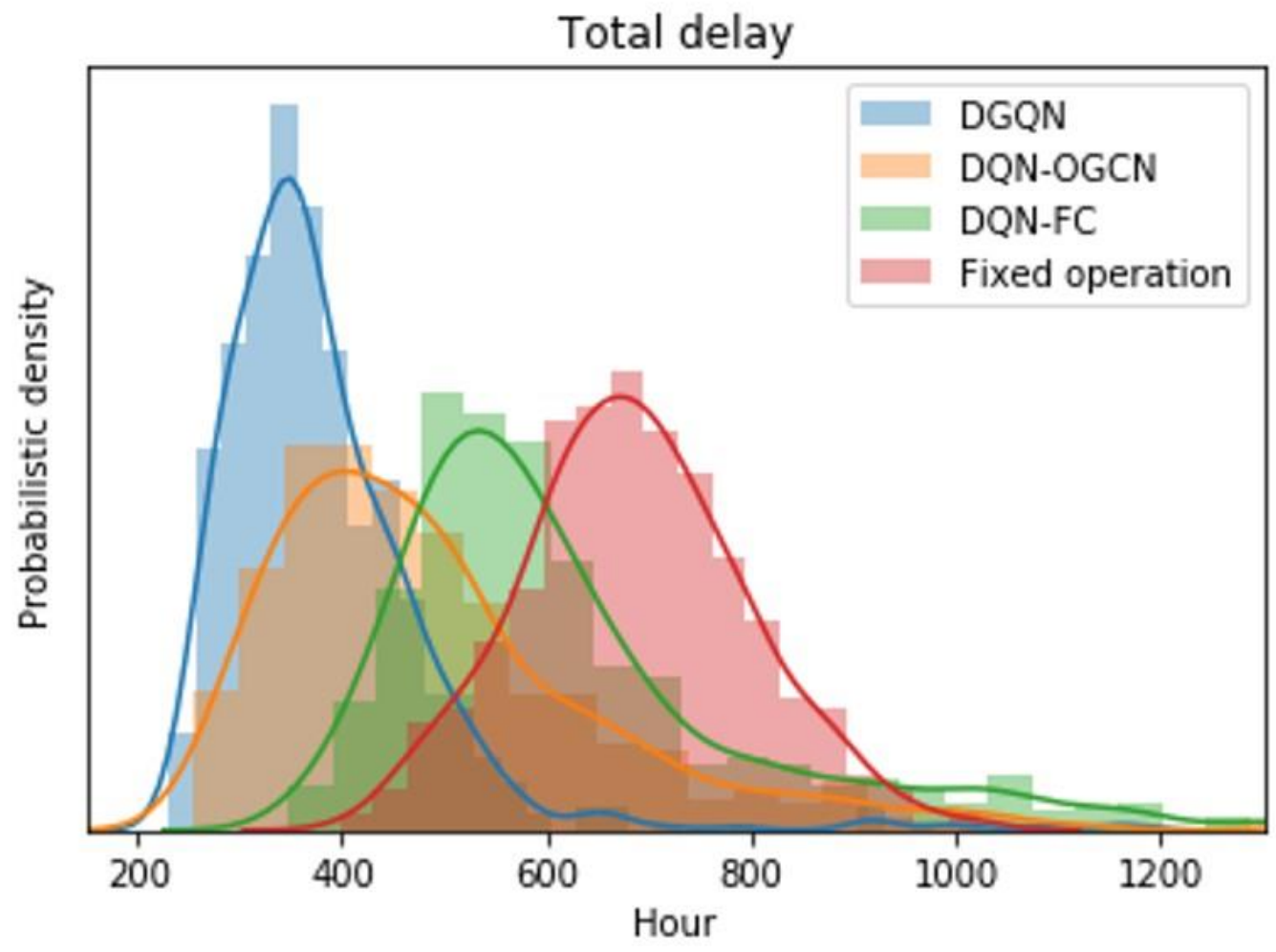

| Total delay (hour) | DGQN | | DQN-OGCN | | DQN-FC | | Baseline (Current fixed operation) | |
|---|---|---|---|---|---|---|---|---|
| | Mean | Standard deviation | Mean | Standard deviation | Mean | Standard deviation | Mean | Standard deviation |
| | 381.9 | 109.8 | 496.8 | 178.1 | 630.3 | 196.6 | 685.2 | 110.3 |

(a) A comparison of model performance based on the total traffic delay

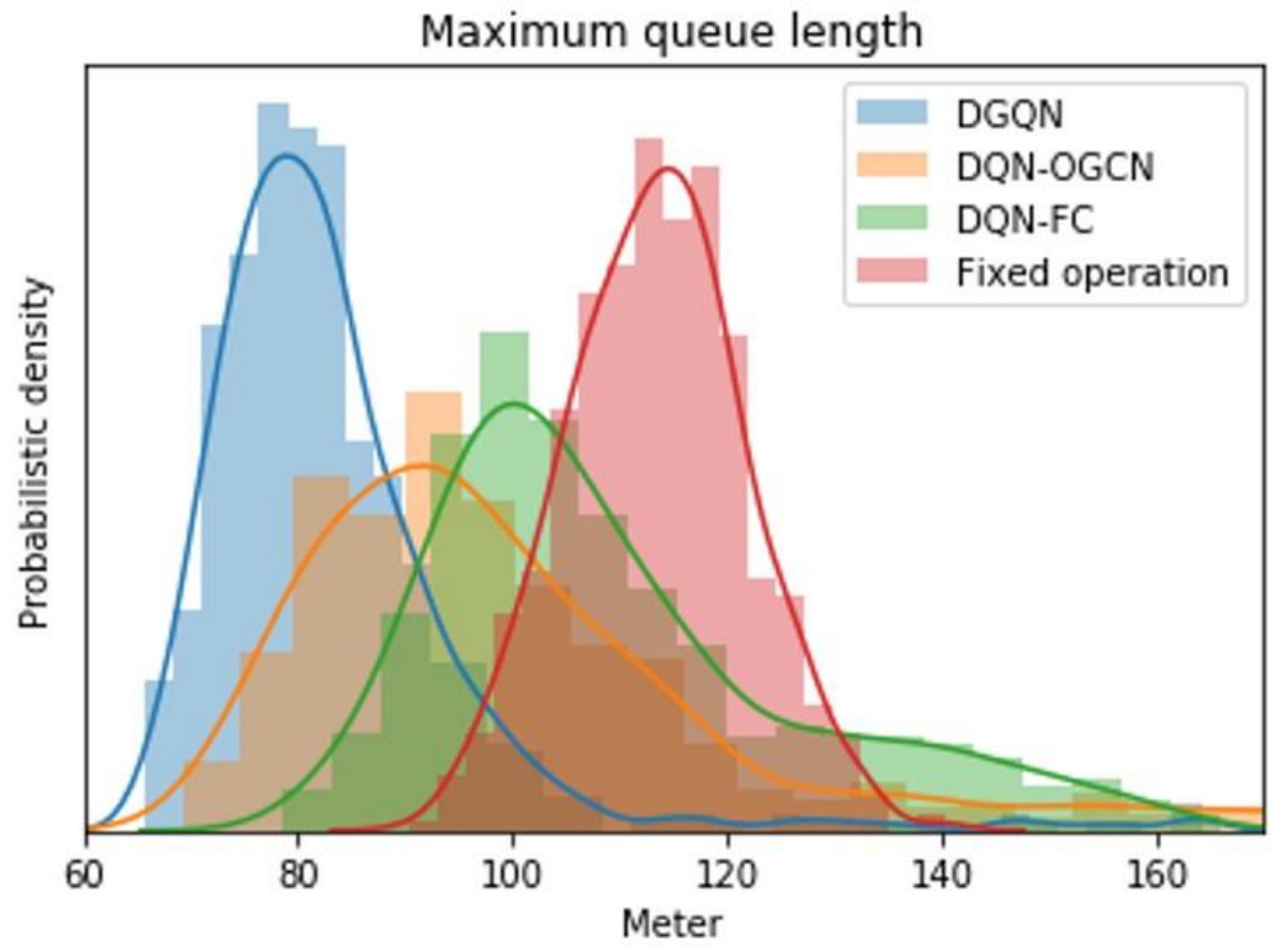

| Maximum queue length (m) | DGQN | | DQN-OGCN | | DQN-FC | | Baseline (Current fixed operation) | |
|---|---|---|---|---|---|---|---|---|
| | Mean | Standard deviation | Mean | Standard deviation | Mean | Standard deviation | Mean | Standard deviation |
| | 83.4 | 13.1 | 101.8 | 24.3 | 109.6 | 17.7 | 113.5 | 8.2 |

(b) A comparison of model performance based on the total traffic delay

Fig. 10. Testing results and comparisons

It should be noted that there were outliers for a DGQN in both performance indices, even though it recorded the best performance on average. This problem is common to all available RL algorithms, since the RL environment of traffic light control is intrinsically nonstationary. The traffic condition within an RL environment is affected by varying the entry traffic volumes and their dynamic route choice patterns, which are totally exogenous and cannot be known in advance. This means that the transition probabilities for a MDP are inconsistent. To overcome this complication, Padakandla and Bhatnagar (2019) proposed a robust method to dynamically differentiate the state regimes of a RL environment. We are attempting to apply this method to traffic signal control on a large scale. In further studies we expect to reduce the variance in traffic delays and queue lengths.

## 8. Conclusions and further studies

The contribution of the present study is three-fold. First, a RL model was devised to accommodate a large action space, and its applicability to real-world traffic signal control on a large scale was verified. Second, a novel graph convolution with a variable adjacency matrix was embedded in a DQN model, and its advantage was substantiated. Last, a practical way to secure the convergence of a large-scale RL model was devised based on asynchronous updating. The advantage of these three aspects was confirmed through simulation experiments that were rigorously set up to mimic the real world.

Nonetheless, we have an important task to fulfill. As far as we could ascertain, there is no RL-based traffic signal controller that is operating in the field. It is very difficult to adapt a RL-based controller to a real-world situation, even after the model has been fully trained on simulated environments. While fine-tuning a RL model in the field, drivers must be resistant against unexpected traffic delays due to traffic signals that a RL controller randomly determines. We are devising a methodology to fine-tune a pre-trained RL model with a large change in states excluded. If successful, a DGQN is expected to be operational in real-world settings.

## Acknowledgement

This research was supported by the Chung-Ang University Research Scholarship Grants in 2019, in part by the National Research Foundation of Korea (NRF) Grant funded by the Korean Government (2018R1A2B200409213), and in part by the Korea Agency for Infrastructure Technology Advancement (KAIA) grant funded by the Ministry of Land, Infrastructure and Transport (Grant 20TLRP-B148659-03).